\documentclass[journal,final,twoside]{IEEEtran}
\usepackage{cite}
\ifCLASSINFOpdf
   \usepackage[pdftex]{graphicx}
   \graphicspath{{./figures/}{./pdf/}{./jpeg/}}
   \DeclareGraphicsExtensions{.pdf,.jpeg,.png}
\else
   \usepackage[dvips]{graphicx}
   \graphicspath{{./eps/}}
   \DeclareGraphicsExtensions{.eps}
\fi
\usepackage{algorithm}
\usepackage{algpseudocode}

\usepackage[cmex10]{amsmath}
\usepackage{amssymb}

\usepackage[caption=false,font=footnotesize]{subfig}
\usepackage{stfloats}

\usepackage{multirow}

\usepackage{footnote}

\usepackage{threeparttable}

\renewcommand{\vec}[1]{\mathbf{#1}}

\newcommand{\mtx}[1]{\mathbf{#1}}

\begin{document}
\title{Optimized Pre-Compensating Compression}

\author{Yehuda Dar, Michael Elad, and Alfred M. Bruckstein % <-this % stops a space
%\thanks{EDICS: }% <-this % stops a space
\\
\thanks{The authors are with the Department of Computer Science, Technion, Israel. E-mail addresses: \{ydar,~elad,~freddy\}@cs.technion.ac.il.}
}


%\markboth{Journal of \LaTeX\ Class Files,~Vol.~6, No.~1, January~2007}%
\markboth{}%
{~}



\maketitle

\begin{abstract}
 
 In imaging systems, following acquisition, an image/video is transmitted or stored and eventually presented to human observers using different and often imperfect display devices. While the resulting quality of the output image may severely be affected by the display, this degradation is usually ignored in the preceding compression. In this paper we model the sub-optimality of the display device as a known degradation operator applied on the decompressed image/video. We assume the use of a standard compression path, and augment it with a suitable pre-processing procedure, providing a compressed signal intended to compensate the degradation without any post-filtering. Our approach originates from an intricate rate-distortion problem, optimizing the modifications to the input image/video for reaching best end-to-end performance.  We address this seemingly computationally intractable problem using the alternating direction method of multipliers (ADMM) approach, leading to a procedure in which a standard compression technique is iteratively applied. We demonstrate the proposed method for adjusting HEVC image/video compression to compensate post-decompression visual effects due to a common type of displays.
 Particularly, we use our method to reduce motion-blur perceived while viewing video on LCD devices.
 The experiments establish our method as a leading approach for preprocessing high bit-rate compression to counterbalance a post-decompression degradation.
 
\end{abstract}


~\\~

% \fi
\IEEEpeerreviewmaketitle

\section{Introduction}
\begin{figure*}[]
	\centering
	\includegraphics[width=0.8\textwidth]{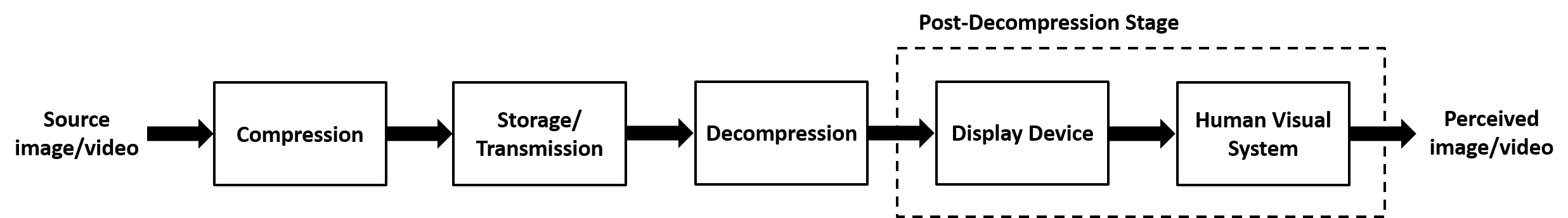}
	\caption{The considered flow of an image/video that is first compressed and finally perceived by a human observer.} 
	\label{Fig:problem_settings_demonstration_high_level}
\end{figure*}

\IEEEPARstart{I}{mage} and video signals have a significant, constantly growing, role in many contemporary applications. 
A fundamental need of image/video applications is to store and/or transmit a digital version of the signal, obeying a bit-budget constraint stemming from the available storage space or the communication channel bandwidth. This bit-budget limitation is managed by lossy compression that produces a compressed representation satisfying the bit-cost constraint at the expense of some distortion in the decompressed signal. 
This systematic flow (see Fig. \ref{Fig:problem_settings_demonstration_high_level}) usually ends with a human user watching the image/video on a display device. Accordingly, the quality of the viewed signal is determined by the compression, the imperfections of the display device, and the human visual system. 

Lossy image and video compression methods trade-off the compressed-form bit-rate with distortion of the decompressed signal. 
Popular compression techniques (e.g., JPEG \cite{wallace1992jpeg}, JPEG2000 \cite{christopoulos2000jpeg2000}, HEVC \cite{sullivan2012overview}) substantially differ in their rate-distortion optimization strategies and the employed image/video models. However, these standard designs ignore other procedures possibly accompanying the compression, thus, may result in sub-optimal rate-distortion performance when considering the complete system.

In this work, we study an intriguing extension of the regular compression problem, where the decompressed signal is degraded by a known linear operator (see Fig. \ref{Fig:problem_settings_demonstration}). Our goal is to compress by considering the squared error between the degraded decompression and the input image. The corresponding rate-distortion optimization has a challenging structure due to the degradation operator involved in the distortion term. We tackle the intricate optimization using the alternating direction method of multipliers (ADMM) approach \cite{boyd2011distributed}, mapping the task to a sequence of easier problems including regular rate-distortion optimizations that are replaced with repeated applications of a standard compression technique. Remarkably, our iterative procedure generically adapts a regular compression method to consider the extended settings involving a post-decompression degradation.

Our approach presented here is a paradigm for managing complicated rate-distortion optimizations associated with sophisticated compression frameworks. 
Specifically, we recently studied the topic of complexity-regularized restoration of an image from its deteriorated version \cite{dar2016image,dar2017restoration}, where the task is an intricate rate-distortion optimization that highly resembles the optimization structure exhibited in this paper for compression purposes. Indeed, we address both problems via iterative optimization procedures, emerging from variable splitting, promoting usage of standard compression techniques. 
In addition, one may assess the vast potential of the proposed compression paradigm for various applications by contemplating the wide use of ADMM  \cite{afonso2010fast,venkatakrishnan2013plug,romano2017little,dar2016postprocessing,dar2016reducing,rond2016poisson} and the similar Half Quadratic Splitting method \cite{geman1995nonlinear,zoran2011learning,dar2016image} for a diversity of challenging signal restoration tasks.

The first part of our experiments considers the adjustment of HEVC image compression to a blur operator degrading the decompressed image. Our results demonstrate the effectiveness of the proposed approach, having superior rate-distortion performance compared to a regular HEVC compression.
Another alternative to accommodate post-decompression degradation is by preceding the compression with a regular deblurring of the input, using the EPLL method \cite{zoran2011learning}. Our method outperforms the EPLL-based approach at high bit-rate compression, reaching impressive average-PSNR (i.e., BD-PSNR \cite{bjontegaard2001calculation}) gains of 2-3 dB. 

As an important application of these ideas, we present a methodology for pre-compression treatment of motion-blur occurring while viewing videos on Liquid Crystal Displays (LCD). 
The prevalent technology of LCD devices relies on a hold-type mechanism, where each frame is constantly displayed until its replacement, resulting in delicate discontinuities of motions. 
The human eye tracks an object based on its smooth motion, trying to fix its location on the retina for a vivid perception.
The smooth eye tracking of discontinuous motion displayed on LCD yields an unsteady positioning on the retina, causing a blurred perception of the moving object.
This blur artifact is amplified for more rapid motions and/or when the video or the display frame-rates are inadequately low, implying too long constant-frame display duration. 
Importantly, motion blur due to the hold-type nature is still an issue of great interest in contemporary evaluations of LCD screens (for examples, see the technical reviews in \cite{rtingsMotionBlurofTVs} and the excellent experimental demonstrations therein) and considered as a crucial drawback of ultra high-definition displays \cite{masia2013survey}.

Straightforward amendments for LCD motion-blur reduce the constant-frame display duration by black-frame insertion \cite{hong2005motion} that causes unwanted eye strains, or by interpolation-based frame-rate up conversion \cite{mishima2004novel,chen2005nonlinearity,dar2015motion} that is computationally intensive and unsuited for complicated motion types.
More sophisticated techniques \cite{klompenhouwer2004motion,har2008lcd,chan2011lcd} counteract the LCD motion blur by a pre-display frame filtering, designed based on blur models of the LCD hold-type behavior and the eye-tracking capability of the human visual system.
These works achieved high PSNR gains using inverse filtering, \cite{klompenhouwer2004motion}, and the Lucy-Richardson deconvolution method \cite{har2008lcd}, however, introduced subjectively annoying noise artifacts that were attenuated in \cite{chan2011lcd} using spatio-temporal smoothness regularization. 

While our application for LCD motion-blur reduction relates to the line of works \cite{klompenhouwer2004motion,har2008lcd,chan2011lcd}, we are the first to address the problem via a pre-compression procedure suggesting computational and accuracy benefits. First, many video content types (e.g., entertainment) are compressed in offline settings rich in computation and time resources, contrasting the regular processing \cite{klompenhouwer2004motion,har2008lcd,chan2011lcd} intended for the display device. 
Accordingly, one can utilize our method in a video-on-demand system designed such that the display types are known and the suitable videos can be delivered to the users.
Second, the blur-compensating filters make use of the current video-motion imperfectly estimated on the available data. While "on-device" methods should practically operate on decompressed frames leading to increased motion-estimation errors (especially at medium/low qualities), our approach uses the pre-compression frames for better motion estimation providing more accurate blur characterization and filtering.
Nicely, our display-blur compensation is, in fact, constrained by the associated video coding procedure acting as a spatio-temporal complexity regularizer preferring smoother or other model-conforming signals costing less bits (see, e.g., in \cite{dar2016image,dar2017restoration}). Consequently, our motion-blur reduction technique provides impressive PSNR gains (with respect to the compression bit-rates) and a pleasing subjective quality.

This paper is organized as follows. In Section \ref{sec:The Proposed Method} we present our method in its general form. In Section \ref{sec:Adjusting HEVC Image Compression to Blurry Decompression} the proposed approach is experimentally studied for adjusting HEVC image compression to balance a post-decompression blur. 
In Section \ref{sec:Application to LCD Motion-Blur Reduction} we employ our method for adapting HEVC video coding to reduce motion blur occurring later on the LCD display. 
Section \ref{sec:Conclusion} concludes this paper.

%The LCD response time, refers to the transition of a pixel value updated to the new frame content, is another cause for blur that has been somewhat alleviated along the years -- yet, still existing in many screens (see, e.g., the current technical reviews in [RTINGS,BLURBUSTER]).

\section{The Proposed Method}
\label{sec:The Proposed Method}

\subsection{The Basic Rate-Distortion Optimization}

\begin{figure*}[]
	\centering
	\includegraphics[width=0.8\textwidth]{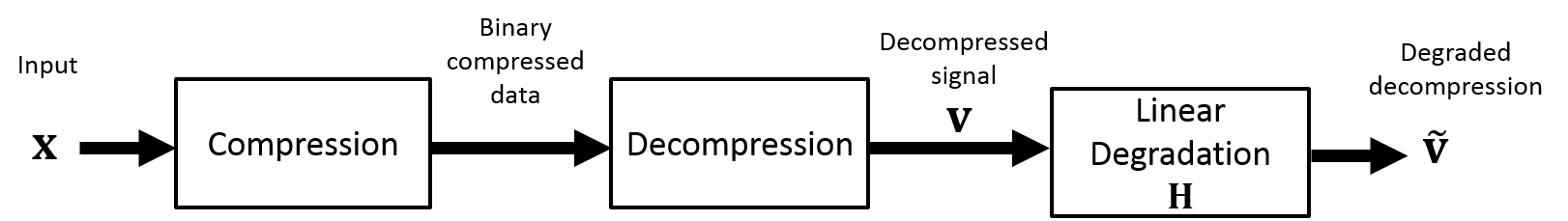}
	\caption{Demonstration of the conceptual problem settings for compression that is oriented to post-decompression degradation.} 
	\label{Fig:problem_settings_demonstration}
\end{figure*}

We develop our method based on the system structure illustrated in Fig. \ref{Fig:problem_settings_demonstration} and explained next. First, an $ N $-dimensional input signal, $ \vec{x} \in \mathbb{R}^N$, goes through a lossy compression procedure resulting in a compressed binary description associated with an approximation of $ \vec{x} $, denoted as $ \vec{v} \in \mathbb{R}^N$, obtained after the decompression stage. However, the reconstruction $ \vec{v}$ is further deteriorated, for instance, due to a sub-optimal display device. We consider here a linear deterioration operator, represented by the $ N\times N $ real-valued matrix $ \mtx{H} $. Then, the degraded decompressed signal is defined as 
\begin{IEEEeqnarray}{rCl}
	\label{eq:degraded decompressed signal - definition}
	\tilde{\vec{v}} \triangleq \mtx{H} \vec{v} , 
\end{IEEEeqnarray}
the outcome of the entire process.

Our goal here is to optimize the compression procedure with respect to the squared error between the input signal $ \vec{x} $ and the degraded decompression $ \tilde{\vec{v}} $, that using (\ref{eq:degraded decompressed signal - definition}) can be expressed as
\begin{IEEEeqnarray}{rCl}
	\label{eq:distortion between input and degradaed decompressed signals}
	D_{H}\left(\vec{x}, \vec{v}\right) \triangleq  \left\| { \vec{x} - \mtx{H} \vec{v} } \right\|_2^2 .
\end{IEEEeqnarray}
Without loss of generality (as will be explained later), we develop our method with respect to a block-based compression design individually operating on blocks of $ N_b $ samples defined by a non-overlapping segmentation of the signal.
We refer to members belonging to the grid of non-overlapping blocks via the set of indices $ \mathcal{B} $.
The block-level compression procedure is modeled as a vector quantizer having a codebook $ \mathcal{C} $, being a finite set of block-reconstruction candidates and their respective variable-length binary codewords.
Specifically, the block-reconstruction $ \vec{c} \in \mathcal{C} $ has a corresponding binary codeword of length $ r\left( \vec{c} \right) $ defining the respective block bit-cost. 
Accordingly, the total bit-cost can be evaluated from the decompressed blocks $ \left\lbrace \vec{v}_i \right\rbrace _{i\in\mathcal{B}} $ as the sum $ \sum\limits_{i\in\mathcal{B}} {r\left( \vec{v}_i \right)} $. 
We define the matrix $ \mtx{P}_i $ as a linear operator extracting the $ i^{th} $ block from the complete signal by the standard multiplication $\mtx{P}_i\vec{v} = \vec{v}_i$. Then, the bit-cost of the entire signal can be expressed as 
\begin{IEEEeqnarray}{rCl}
	\label{eq:total bit-cost}
	R\left(\vec{v}\right) =  \sum\limits_{i\in\mathcal{B}} {r\left( \mtx{P}_i\vec{v} \right)} .
\end{IEEEeqnarray}

We use the quantities defined in (\ref{eq:distortion between input and degradaed decompressed signals})-(\ref{eq:total bit-cost}) to formulate the rate-distortion optimization in the unconstrained Lagrangian form: 
\begin{IEEEeqnarray}{rCl}
	\label{eq:degradation-oriented rate-distortion optimization}
	\hat{\vec{v}} = \mathop {{\text{argmin}}}\limits_{\vec{v}\in\mathcal{C_{\mathcal{B}}}} ~ \left\| { \vec{x} - \mtx{H} \vec{v} } \right\|_2^2 + \lambda \sum\limits_{i\in\mathcal{B}} {r\left( \mtx{P}_i\vec{v} \right)}
\end{IEEEeqnarray}
where $ \lambda \ge 0 $ is the Lagrange multiplier associated with some total bit-cost constraint, and $\hat{\vec{v}}$ is the optimal decompressed signal among the candidates available in the effective full-signal codebook:
\begin{IEEEeqnarray}{rCl}
	\label{eq:full signal codebook}
	\mathcal{C}_{\mathcal{B}} = \left\lbrace {\vec{c} ~~ \Big|{~~ {\vec{c} = \mathop \sum\limits_{i\in\mathcal{B}}{\mtx{P}_i^T \vec{c}_i}, ~~  {\{\vec{c}_i\}_{i\in\mathcal{B}} \in \mathcal{C}}} } }\right\rbrace
\end{IEEEeqnarray}
where the linear operator $ \mtx{P}_i^T $ places a block in the $ i^{th} $ block location in a full-signal layout.
One should note that, for an arbitrarily structured $ \mtx{H} $, the optimization (\ref{eq:degradation-oriented rate-distortion optimization}) is difficult to solve since it does not allow the commonly used block-based treatment (for examples, see its various forms in the fundamental studies on operational rate-distortion optimization \cite{shoham1988efficient,ortega1998rate,sullivan1998rate} and also in recent works \cite{li2014lambda,li2017optimal}).

\subsection{Practical Iterative Procedure}

The structural complication of the rate-distortion optimization (\ref{eq:degradation-oriented rate-distortion optimization}) is facilitated using the ADMM strategy \cite{boyd2011distributed} as explained next. 
Initially, we define the auxiliary variable $ \vec{z} \in \mathbb{R}^N $ letting us to reformulate the problem (\ref{eq:degradation-oriented rate-distortion optimization}) into
\begin{IEEEeqnarray}{rCl}
	\label{eq:degradation-oriented rate-distortion optimization - with splitting - constrained form}
		\left( \hat{ \vec{v}}, \hat{\vec{z}} \right) = \mathop {{\text{argmin}}}\limits_{{\vec{v}\in\mathcal{C_{\mathcal{B}}}},~ {\vec{z}}\in\mathbb{R}^N } \left\| { \vec{x} - \mtx{H} \vec{z} } \right\|_2^2 + 
		\lambda  \mathop\sum\limits_{i\in \mathcal{B}} {{r}( \mtx{P}_i\vec{v} )} \\ 
		\text{s.t. }~~~~~ \vec{z} = \vec{v} . ~~~~~~~~~~~~~~~~~~~~~~~~~~~
\end{IEEEeqnarray}
Then, considering (\ref{eq:degradation-oriented rate-distortion optimization - with splitting - constrained form}) via its augmented Lagrangian (in its scaled version \cite[Ch. 2]{boyd2011distributed}) leads to an iterative procedure, where the $ t^{th} $  iteration is 
\begin{IEEEeqnarray}{rCl}
	\label{eq:degradation-oriented rate-distortion optimization - with augmented Lagrangian}
	&& \left( \hat{ \vec{v}}^{(t)}, \hat{\vec{z}}^{(t)} \right) = 
	\\ \nonumber
	&& \mathop {{\text{argmin}}}\limits_{{\vec{v}\in\mathcal{C_{\mathcal{B}}}},~ {\vec{z}}\in\mathbb{R}^N } \left\| { \vec{x} - \mtx{H} \vec{z} } \right\|_2^2   + \lambda  \sum\limits_{i\in \mathcal{B}} r(\mtx{P}_i\vec{v}) + \frac{\beta}{2}{\left\| { \vec{v} - \vec{z} + \vec{u}^{(t)} } \right\|_2^2} 
	\\ 
	&& \vec{u}^{(t+1)} = \vec{u}^{(t)} + \left( \hat{ \vec{v}}^{(t)} - \hat{\vec{z}}^{(t)} \right),
\end{IEEEeqnarray}
where $ \vec{u}^{(t)} \in \mathbb{R}^N$ is the scaled dual variable and $ \beta $ is an auxiliary parameter originating at the Lagrangian.

Since each of the optimization variables in (\ref{eq:degradation-oriented rate-distortion optimization - with augmented Lagrangian}) participates only in two of the three terms in the cost function and, therefore, one iteration of alternating minimization provides us the ADMM form that iterates over the following manageable optimizations: 
\begin{IEEEeqnarray}{rCl}
	\label{eq:degradation-oriented rate-distortion optimization - iterative solution - compression}
	\hat{\vec{v}}^{(t)} = \mathop {\text{argmin}}\limits_{\vec{v}\in\mathcal{C_{\mathcal{B}}}} \frac{\beta}{2}{\left\| { \tilde{ \vec{z}}^{(t)} - \vec{v} } \right\|_2^2}  + \lambda  \sum\limits_{i\in \mathcal{B}} r(\mtx{P}_i\vec{v})
\end{IEEEeqnarray}
\begin{IEEEeqnarray}{rCl}
	\label{eq:degradation-oriented rate-distortion optimization - iterative solution - inversion}
	\hat{ \vec{z}}^{(t)} = \mathop {\text{argmin}}\limits_{{\vec{z}}\in\mathbb{R}^N } \left\| { \vec{x} - \mtx{H} \vec{z} } \right\|_2^2 + \frac{\beta}{2}{\left\| {  \vec{z} - \tilde{ \vec{v}}^{(t)} } \right\|_2^2}
\end{IEEEeqnarray}
\begin{IEEEeqnarray}{rCl}
	\label{eq:degradation-oriented rate-distortion optimization - iterative solution - u update}
	\vec{u}^{(t+1)} = \vec{u}^{(t)} + \left( \hat{ \vec{v}}^{(t)} - \hat{\vec{z}}^{(t)} \right).
\end{IEEEeqnarray}
where $ \tilde{ \vec{z}}^{(t)} = \hat{\vec{z}}^{(t-1)} - \vec{u}^{(t)} $ and $ \tilde{ \vec{v}}^{(t)} = \hat{\vec{v}}^{(t)} + \vec{u}^{(t)} $.
The analytic solution of the second-stage problem in (\ref{eq:degradation-oriented rate-distortion optimization - iterative solution - inversion}) is 
\begin{IEEEeqnarray}{rCl}
	\label{eq:degradation-oriented rate-distortion optimization - iterative solution - inversion - analytic form}
	\hat{ \vec{z}}^{(t)} = \left(  \mtx{H}^{T}\mtx{H} + \frac{\beta}{2} \mtx{I}  \right)^{-1} \left( \mtx{H}^T \vec{x} + \frac{\beta}{2} \tilde{ \vec{v}}^{(t)}  \right),  
\end{IEEEeqnarray}
thus, exhibiting optimization (\ref{eq:degradation-oriented rate-distortion optimization - iterative solution - inversion}) as a weighted averaging operation.

Importantly, the first stage (\ref{eq:degradation-oriented rate-distortion optimization - iterative solution - compression}) is a rate-distortion optimization compatible with a block-based treatment and considering the regular squared-error metric for the compression of $ \tilde{ \vec{z}}^{(t)} $, obtained in the second stage of the former iteration. Moreover, this full-image rate-distortion optimization is done for a Lagrange multiplier of value $ \tilde{\lambda} = \frac{2\lambda}{\beta} $.
We denote the compression-decompression procedure associated with (\ref{eq:degradation-oriented rate-distortion optimization - iterative solution - compression}) as 
\begin{IEEEeqnarray}{rCl}
	\label{eq:compression-decompression function}
	\hat{\vec{v}}^{(t)}=CompressDecompress_{\lambda} \left( \tilde{\vec{z}}^{(t)}  \right) .
\end{IEEEeqnarray}
We further suggest using a standard compression method as the compression-decompression operator (\ref{eq:compression-decompression function}). While many compression methods do not follow the exact rate-distortion optimization we got in our mathematical development (\ref{eq:degradation-oriented rate-distortion optimization - iterative solution - compression}), we still suggest using such techniques as replacements for (\ref{eq:degradation-oriented rate-distortion optimization - iterative solution - compression}). 
Additionally, since various compression methods do not rely on Lagrangian optimization, their operating parameters may differ (for example, quality parameters, compression ratios, or output bit-rates). 
Accordingly, we present the suggested algorithm with respect to a general compression procedure that its output bit-cost is directly or indirectly affected by a parameter denoted as $ \theta $. This generalization is used in Algorithm \ref{Algorithm:Proposed Method}.

\begin{algorithm}
	\caption{Proposed Method: Compression Adjusted to Post-Decompression Degradation}
	\label{Algorithm:Proposed Method}
	\begin{algorithmic}[1]
		\State Inputs: $ \vec{x} $, $ \theta $, $ \beta $.
		\State  Initialize $ {\hat{\vec{z}}}^{(0)} = \vec{x} $ , $\vec{u}^{(1)} = \vec{0}$.
		\State $t = 1$
		\Repeat 
		
		\State $ \tilde{ \vec{z}}^{(t)} = \hat{\vec{z}}^{(t-1)} - \vec{u}^{(t)} $
		\State $ \hat{\vec{v}}^{(t)} = {CompressDecompress}_{\theta}\left( \tilde{\vec{z}}^{(t)} \right). $
		\vspace{0.05in}
		
		\State $ \tilde{ \vec{v}}^{(t)} = \hat{\vec{v}}^{(t)} + \vec{u}^{(t)} $
		\State $\hat{ \vec{z}}^{(t)} = \left(  \mtx{H}^{T}\mtx{H} + \frac{\beta}{2} \mtx{I}  \right)^{-1} \left( \mtx{H}^T \vec{x} + \frac{\beta}{2} \tilde{\vec{v}}^{(t)}   \right)$
		
		\State $\vec{u}^{(t+1)} = \vec{u}^{(t)} + \left( \hat{ \vec{v}}^{(t)} - \hat{\vec{z}}^{(t)} \right)$
		
		\State $ t \gets t + 1$
		\Until{stopping criterion is satisfied}
		\State Output: Binary compressed data obtained in the last application of Stage 6.
	\end{algorithmic}
\end{algorithm}

The replacement of the rate-distortion optimization formulation in (\ref{eq:degradation-oriented rate-distortion optimization - iterative solution - compression}) with a standard compression-decompression process (\ref{eq:compression-decompression function}) is motivated by a similar development step used in the Plug-and-Play Priors method \cite{venkatakrishnan2013plug} for image restoration, where an optimization stage corresponding to a Maximum A-Posteriori (MAP) Gaussian denoising problem is replaced with the application of an existing denoiser (such as BM3D \cite{dabov2007image}).
In both cases (ours and in \cite{venkatakrishnan2013plug}), the application of an arbitrary compression/denoising method means that the convexity of the optimization problem cannot be guaranteed and, therefore, in some cases the optimization may not converge. Accordingly, the implementations we present in Sections \ref{sec:Adjusting HEVC Image Compression to Blurry Decompression}-\ref{sec:Application to LCD Motion-Blur Reduction} include a divergence detection mechanism as part of the stopping criterion, this feature is explained later in this paper. Studying the convergence/divergence properties of our framework (in the spirit of the analysis in \cite{chan2017plug}) is left to a future work.

\begin{figure}[]
	\centering
	\includegraphics[width=0.45\textwidth]{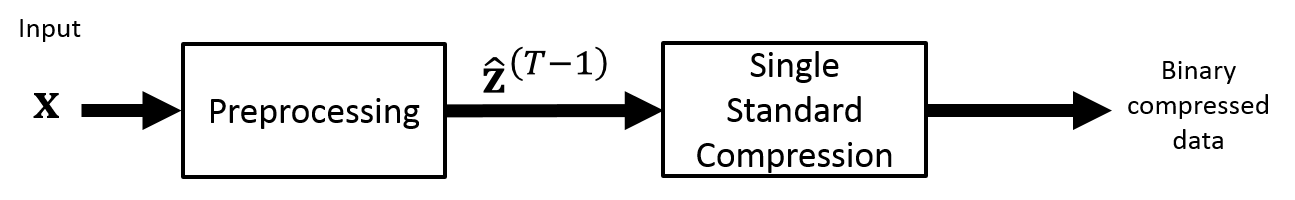}
	\caption{Interpretation of the proposed compression method as a preprocessing stage followed by a single standard compression. The demonstration here assumes that our procedure runs $ T $ iterations.} 
	\label{Fig:proposed_method_preprocessing_interpretation}
\end{figure}

We can further interpret the proposed iterative compression approach as a preprocessing stage coupled with a single standard compression, being the one applied in the last iteration as the determining stage outputing the compressed binary data (see Fig. \ref{Fig:proposed_method_preprocessing_interpretation}). Remarkably, our compression output is compatible with a standard decompression process. 
The overall quality improvement suggested by our method obviously entails an increased computational cost that, nevertheless, is distributed between the encoder and the decoder stages in the following attractive structure: the decoder complexity remains as in the standard form, while the encoder has the increased computational load of repeatedly applying standard compressions (\ref{eq:compression-decompression function}) and the $ \ell_2 $-constrained deconvolutions (\ref{eq:degradation-oriented rate-distortion optimization - iterative solution - inversion - analytic form}). 
This system layout is beneficial for applications where the compression can be carried out offline in environments rich in computational and time resources, whereas the decompression on the display devices should be of a low computational cost due to run-time and energy-consumption limitations.

\section{Adjusting HEVC Image Compression to Blurry Decompression}
\label{sec:Adjusting HEVC Image Compression to Blurry Decompression}

In this section we demonstrate our approach for adapting HEVC's still-image compression (in its version implemented in the BPG image format \cite{hevc_software_bpg}) to a blur deteriorating the decompressed image.
The experiment goal here is to study our method with respect to alternative processing strategies.
Our ideal settings here, considering a known Gaussian blur kernel, serve as a preliminary stage to the intricate application presented in Section \ref{sec:Application to LCD Motion-Blur Reduction}.
% where the motion blur kernel varies with the video content and is imperfectly estimated.
The degradation operator $ \mtx{H} $ is associated with a Gaussian blur kernel of standard deviation 0.6 and $ 15\times 15 $ pixels size. We consider a shift-invariant degradation, thus, efficiently degrade an image using a two-dimensional convolution with the blur kernel. 

The Peak Signal-to-Noise Ratio (PSNR) is defined here based on the squared-error distortion (\ref{eq:distortion between input and degradaed decompressed signals}) and can be written as 
\begin{IEEEeqnarray}{rCl}
	\label{eq:PSNR definition}
	PSNR = 10 \log_{10} \left( \frac{P^2}{\frac{1}{N}\left\| { \vec{x} - \tilde{\vec{v}} } \right\|_2^2} \right)
\end{IEEEeqnarray}
where $\vec{x}$ and $\tilde{\vec{v}}$ are the input and the degraded decompressed signals, respectively, and $ N $ is the signal dimension. The maximal value attainable by the examined signals is denoted as $ P $ that, e.g., equals to $255$ for grayscale images with pixel values in the range $ [0,255] $.
In the PSNR computation we ignore margins of 35 pixels along the borders of the image to exclude effects of specific boundary conditions used in the applied convolutions.

\begin{figure*}[]
	\centering
	{\subfloat[Cards]{\label{fig:cards_a_300x300.png_RD_curves}\includegraphics[width=0.31\textwidth]{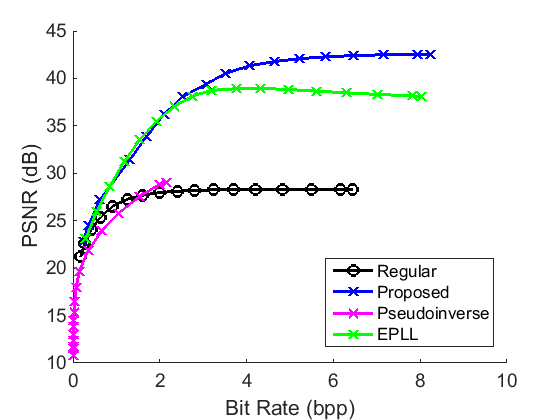}}}~
	{\subfloat[Tree]{\label{fig:ucid_Tree_RD_curves}\includegraphics[width=0.31\textwidth]{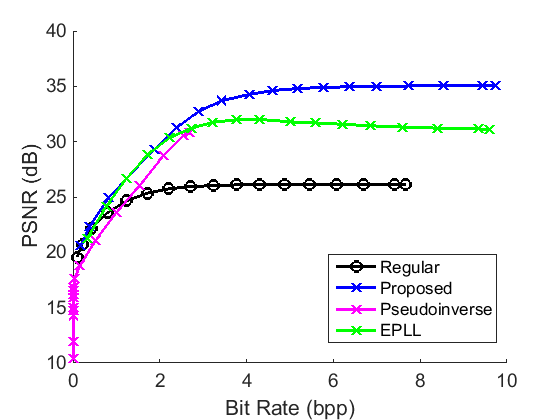}}}~
	{\subfloat[Bears]{\label{fig:berkely_bears.jpg_RD_curves}\includegraphics[width=0.31\textwidth]{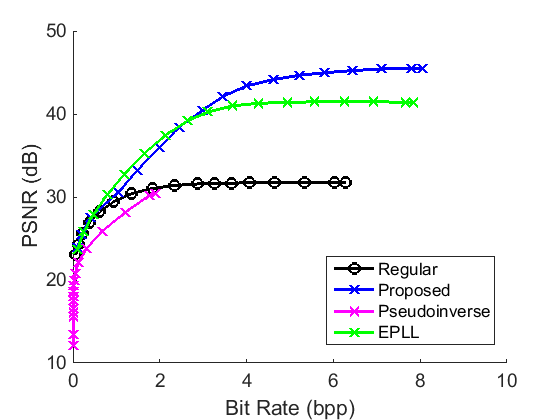}}}
	\caption{PSNR-bitrate curves comparing our approach to competing methods for three grayscale images (see also Table \ref{table:Experiments - blur - Average PSNR and Bit-Rate comparison}). The post-decompression deterioration is a $ 15\times 15 $ Gaussian blur kernel (standard deviation 0.6).} 
	\label{Fig:Experiments - blur - RD curves comparison}
\end{figure*}

\begin{figure*}[]
	\centering
	{\subfloat[TESTIMAGES]{\label{fig:samling_dataset_examples}\includegraphics[height=0.15\textwidth]{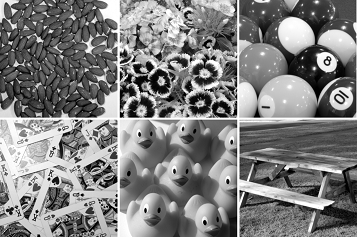}}}~
	{\subfloat[UCID]{\label{fig:UCID_dataset_examples}\includegraphics[height=0.15\textwidth]{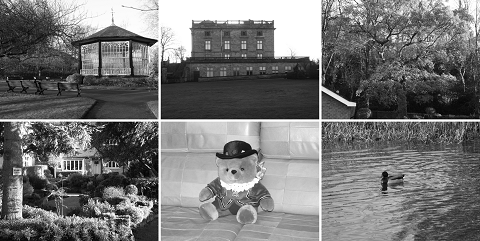}}}~
	{\subfloat[Berkeley]{\label{fig:berkeley_dataset_examples}\includegraphics[height=0.15\textwidth]{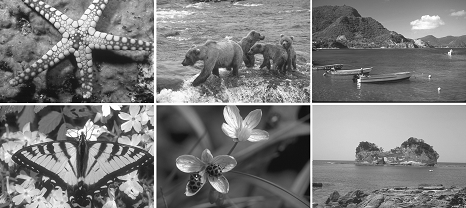}}}
	\caption{The images used in the experiments.} 
	\label{Fig:Experiments - blur - image dataset examples}
\end{figure*}

\begin{table*} []
	\caption{~~~~~~~~~~~~~Image Compression Considering Post-Decompression Deterioration of a Gaussian Blur\newline Average PSNR Gains (measured using BD-PSNR)}
	\renewcommand{\arraystretch}{1.1}
	\label{table:Experiments - blur - Average PSNR and Bit-Rate comparison}
	\centering
	\begin{tabular}{|c|c||c|c||c|c|}
		\hline
		\multirow{2}{*}{\bfseries \shortstack{Dataset}}  & \multirow{2}{*}{\bfseries \shortstack{Image}}  & \multicolumn{2}{|c|}{\bfseries \shortstack{All Bit-Rates}} &  \multicolumn{2}{|c|}{\bfseries\shortstack{High Bit-Rates}} \\
		\cline{3-6}
		& & Proposed over Regular & Proposed over EPLL & Proposed over Regular & Proposed over EPLL \\
		\hline\hline				                                       
		TESTIMAGES 300x300 & Almonds  & 5.49 & 0.22 & 14.59 & 3.22 \\
		\cline{2-6}	
		& Flowers  & 5.57 & 0.41 & 14.96 & 3.21 \\
		\cline{2-6}	
		& Billiard balls  & 5.57 & -0.08 & 12.60 & 2.00 \\
		\cline{2-6}	
		& Cards  & 6.08 & 0.90 & 13.55 & 3.67 \\
		\cline{2-6}	
		& Duck toys  & 3.62 & -0.62 & 10.42 & 1.08 \\
		\cline{2-6}	
		& Garden table  & 3.78 & 0.29 & 12.61 & 2.69 \\										
		\hline\hline
		UCID 384x512 & Garden house & 3.21 & 1.14 & 7.48 & 3.39 \\
		\cline{2-6}	
		& House and lawn  & 2.41 & 0.50 & 6.41 & 2.97 \\
		\cline{2-6}	
		& Tree  & 3.57 & 1.16 & 8.79 & 3.64 \\
		\cline{2-6}	
		& Garden  & 3.16 & 1.49 & 6.55 & 3.32 \\
		\cline{2-6}	
		& Teddy bear  & 3.65 & -0.39 & 10.99 & 2.37 \\
		\cline{2-6}	
		& Duck  & 5.25 & 0.06 & 15.64 & 2.62 \\
		\hline\hline				                                       
		Berkeley 481x321  & Starfish  & 4.72 & -0.36 & 15.42 & 2.02 \\
		\cline{2-6}	
		& Bears & 3.78 & 0.42 & 12.69 & 3.46 \\
		\cline{2-6}
		& Boats & 3.29 & 0.43 & 8.94 & 2.65 \\
		\cline{2-6}
		& Butterfly & 4.76 & 0.42 & 11.80 & 3.33 \\
		\cline{2-6}
		& Flower and Bugs & 4.18 & 0.02 & 11.00 & 2.39 \\
		\cline{2-6}
		& Sea & 4.58 & 0.21 & 12.87 & 2.49 \\
		\hline\hline		                                       
		\hline		                                               
	\end{tabular}
\end{table*}

\subsection{Competing Methods} 

We compare our approach to three competing strategies also considering HEVC image compression. In Figure \ref{Fig:Experiments - blur - RD curves comparison} we compare the rate-distortion curves of the various methods corresponding to operating their HEVC component using quantization parameter (QP) values between 1 to 49 in jumps of 3. The examined compression procedures are:

\subsubsection{Regular compression without any pre/post processing} This is the baseline approach where a regular compression-decompression application is followed by deterioration. Obviously, since this procedure ignores the degradation, it is the cheapest in computations and provides an inferior performance (see the solid-line black curves in Fig. \ref{Fig:Experiments - blur - RD curves comparison}).

%\subsubsection{Pseudoinverse filtering of the decompressed image before degradation} 
%This approach processes the decompressed image before its degradation (i.e., in terms of Fig. \ref{Fig:problem_settings_demonstration} the signal $ \vec{v} $ is filtered and the outcome being degraded). Evidently, this representative method deviates from our goal to conduct a pre-compression processing due to computational and/or other limitations in the post-decompression stage. We consider here an ideal pseudoinverse filter matched to the subsequent degradation operator $ \mtx{H} $ (accurately known via an oracle capability), thus, perfectly canceling the degradation except to loss of signal components belonging to $ \mtx{H} $'s null space. 
%Since this approach relies on post-decompression processing and numerical ideal abilities, it is not a fair competitor to the other methods presented -- thus, brought here for the purpose of performance reference. Indeed, it provides the best rate-distortion trade-offs at medium and high bit-rates, and performs similar to other methods at low bit-rates (see the dashed-line magenta curves in Figs. ??-??).

\subsubsection{Pre-compression pseudoinverse filtering of the input image} 
We consider here an ideal pseudoinverse filter, matched to the known degradation operator $ \mtx{H} $, employed as a pre-compression filter. 
The numerical crudity of the pseudoinverse filter yields a very large dynamic range of pixel values, 
hence, requiring shifting and scaling before the compression and the inverse adaptations after decompression (before degradation). Since the pseudoinverse filtered image is far from obeying natural-image characteristics (e.g., smoothness), it is inefficiently compressed by a standard compression technique. This drawback results in performance inferior even to regular compression without any processing (see Fig. \ref{Fig:Experiments - blur - RD curves comparison}). 
Moreover, the unusual signals provided by the pseudoinverse filter can be compressed using HEVC to a limited range of bit-rates, for examples, observe the rightmost working-points of the magenta curves obtained using HEVC compression at the very high quality corresponding to $QP=1$.
This exemplary approach exhibits the challenges in pre-compression processing.

\subsubsection{Pre-compression filtering via the Expected Patch Log Likelihood (EPLL) method} 
We define the main competing method to employ a pre-compression filtering in the form of the EPLL deblurring method relying on a Gaussian Mixture Model (GMM) prior learned for natural images (see \cite{zoran2011learning}). Indeed, the processed image conforms with natural-image attributes, thus, efficiently compressed by HEVC leading to a good rate-distortion performance considering the degraded decompressed image (see the solid-line green curves in Fig. \ref{Fig:Experiments - blur - RD curves comparison}). In the EPLL experiments we used the implementation published by the authors of \cite{zoran2011learning} with parameters we found to improve the rate-distortion performance considered here.

\subsection{Our Method: Experiment Settings} 
\label{subsec: Image experiments - our method}

We now turn to evaluate our method with respect to the above three reference techniques. In the implementation of Algorithm \ref{Algorithm:Proposed Method} we set $ \beta $ to a value depending on the specific quantization parameter (QP) given to the HEVC compression:
\begin{IEEEeqnarray}{rCl}
	\label{eq:beta in image experiments}
	\beta = \left\{ {\begin{array}{*{20}{c}}
			0.03&{,\text{for~~} 0 \le QP \le 20 } \\ 
			0.05&{,\text{for~~} 20 < QP \le 30 } \\  
			0.10&{,\text{for~~} 30 < QP \le 40 } \\  
			0.35&{,\text{for~~} 40 < QP \le 45 } \\  
			0.45&{,\text{for~~} 45 < QP \le 51 } \\  
		\end{array}} \right.  	
\end{IEEEeqnarray}
Recall that HEVC QP values are integers between 0 to 51, where a lower value yields a higher quality.
The stopping criterion was defined to a maximal number of 40 iterations or to end earlier when $ \hat{\vec{v}}^{(t)} $ and $ \hat{ \vec{z}}^{(t)} $ are detected to converge or diverge. The convergence/divergence detection relies on the total absolute difference between $  \hat{\vec{v}}^{(t)} $ and $ \hat{ \vec{z}}^{(t)} $ in each iteration, namely, $ w^{(t)} \triangleq \left\| {  \hat{\vec{v}}^{(t)} - \hat{ \vec{z}}^{(t)} } \right\|_1 $. Accordingly, we determine convergence when 
$ \lvert w^{(t)} - w^{(t-1)} \rvert < 0.2  $ for three consecutive iterations. Divergence is identified when $ w^{(t)} - w^{(t-1)} > 50  $ and, in that case, the algorithm output is taken from the preceding iteration.  
Note that the threshold values given here for convergence/divergence detection depend on the signal dimension and the typical value-range (the thresholds specified here are for signal with values in the range [0,1]).

\begin{figure*}[]
	\centering
	{\subfloat[Input]{\label{fig:starfish_original}\includegraphics[width=0.28\textwidth]{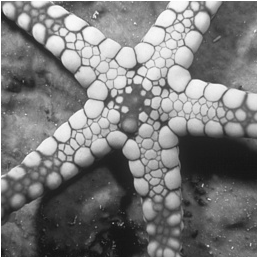}}}~~
	{\subfloat[Regular Decompression (5.061 bpp)]{\label{fig:starfish_blur_regular_decompression_before_degradation_5_061_bpp}\includegraphics[width=0.28\textwidth]{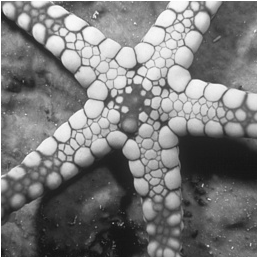}}}~~
	{\subfloat[Regular Degraded Decompression (34.33 dB)]{\label{fig:starfish_blur_regular_degraded_decompression_5_061__bpp_psnr_34_33_dB}\includegraphics[width=0.28\textwidth]{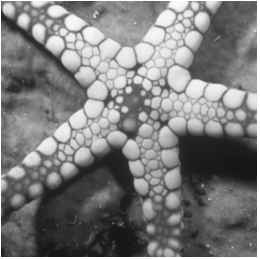}}}
	\\
	{\subfloat[Our: Input to Last Iteration Compression]{\label{fig:starfish_blur_our_input_to_last_iteration_compression_4_296_bpp}\includegraphics[width=0.28\textwidth]{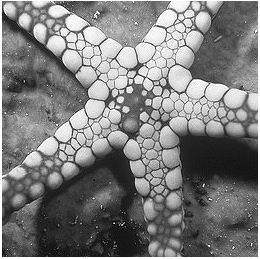}}}~~
	{\subfloat[Our: Decompression (4.296 bpp)]{\label{fig:starfish_blur_our_decompression_before_degradation_4_296_bpp}\includegraphics[width=0.28\textwidth]{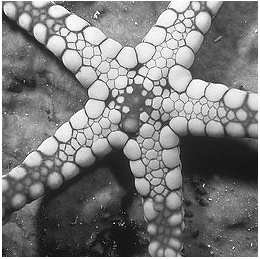}}}~~
	{\subfloat[Our: Degraded Decompression (49.58 dB)]{\label{fig:starfish_blur_our_degraded_decompression_4_296_bpp_psnr_49_58_dB}\includegraphics[width=0.28\textwidth]{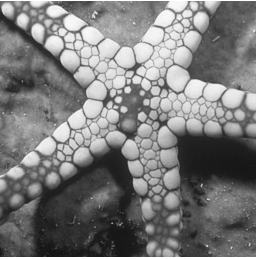}}}
	\caption{Demonstrating the intermediate and resultant images of the regular and the proposed compression methods for a post-decompression deterioration of a Gaussian blur kernel. The image is 'Starfish' (a segment of $ 256\times 256 $ pixels is shown).} 
	\label{Fig:Experiments - blur - starfish}
\end{figure*}

The rate-distortion curves of our method, presented in Fig. \ref{Fig:Experiments - blur - RD curves comparison} as the blue solid-lines, outperform the other pre-compression techniques at the high bit-rate range. The PSNR gains at high bit-rates are significant (note the wide PSNR range of the graphs that may visually mislead), reaching improvements of several dBs.
%Performance improvement at high bit-rates, associated with high reconstruction qualities, is important for applications presenting entertainment content. Nicely, these applications are also suitable for pre-compression processing due to a possible offline compression.
These impressive PSNR gains at high bit-rates were further established by examining 18 images (see Fig. \ref{Fig:Experiments - blur - image dataset examples} and Table \ref{table:Experiments - blur - Average PSNR and Bit-Rate comparison}) collected from three different datasets (TESTIMAGES \cite{asuni2014testimages}, UCID \cite{schaefer2003ucid}, and Berkeley \cite{martin2001database}). The comparison in Table \ref{table:Experiments - blur - Average PSNR and Bit-Rate comparison} considers the average PSNR difference between performance curves (e.g., see Fig. \ref{Fig:Experiments - blur - RD curves comparison}) of the proposed, the EPLL-based, and the regular methods. The average PSNR differences between curves were calculated using the BD-PSNR metric \cite{bjontegaard2001calculation,BDPSNR_Matlab}, for the entire bit-rate range (i.e., the complete curves generated for QP values between 1 to 49 in jumps of 3) and for curve segments corresponding to high bit-rates (defined by QP values 1, 7, 13 and 19).

In Figure \ref{Fig:Experiments - blur - starfish} we present visual results for the 'Starfish' image. First, we examine the regular compression procedure where the input image (Fig. \ref{fig:starfish_original}) is compressed using HEVC at a bit-rate of 5.061 bits per pixel (bpp), leading to the decompressed image in Fig. \ref{fig:starfish_blur_regular_decompression_before_degradation_5_061_bpp} (note that this is the pre-degradation image). Then,  obviously, the post-degradation decompressed image (Fig. \ref{fig:starfish_blur_regular_degraded_decompression_5_061__bpp_psnr_34_33_dB}) suffers from tremendous blur affecting also the PSNR (measured with respect to the precompression image).
In contrast, our approach processes the input image such that the compression in the last iteration gets a sharpened version (see Fig. \ref{fig:starfish_blur_our_input_to_last_iteration_compression_4_296_bpp}) adjusted to the specifically known blur operator, then, the compressed image at bit-rate 4.296 bpp leads to a degraded decompression with moderate blur effects (Fig. \ref{fig:starfish_blur_our_degraded_decompression_4_296_bpp_psnr_49_58_dB}) and PSNR improvement of 15.25 dB with respect to the regular compression at a higher bit-rate.

The experiments show that the current implementation of our approach is significantly better than the considered alternatives for compression at the high bit-rate range. We consider this behavior to emerge from the following two facts. First, the details of the pre-compression processing are preserved better when high bit-rate compression is applied. Second, at high bit-rate compression the employed quantization is finer, thus, the discrete optimization problem imitates more closely an optimization over a continuous domain – which is more suitable for the ADMM optimization technique.

\section{Application to LCD Motion-Blur Reduction}
\label{sec:Application to LCD Motion-Blur Reduction}

\subsection{The LCD Motion Blur and Its Modeling}
\label{subsec:The LCD Motion Blur Model}

A prominent type of post-decompression degradations is defined as the inevitable artifacts arising due to various display device technologies.
For instance, the formerly prevalent Cathode Ray Tube (CRT) displays employ an impulse-type mechanism where video frames are instantaneously presented, producing a good perceptual motion-continuity and unpleasing flickering artifacts. Here we focus on the current Liquid Crystal Display (LCD) technology, the ultimate successor of CRT, being a hold-type display where each frame is constantly presented for a duration of $ ({frame~rate})^{-1}  $ seconds, referred to as the hold time (e.g., for a rate of 60 frames per second the hold time is 16.6 milliseconds). While this hold-type architecture is flickering free, it suffers from a non-smooth presentation of motions that cause blur in the image perceived by the viewer.  
Specifically, the human eye pursues constant motion of an object to fix its image location on the retina for a detailed perception. While motion presented on an LCD device has delicate discontinuities, the eye still tracks it as if it was continuous and, thus, suffers from corresponding spatial displacements on the retina that blur the perceived image.

Additional LCD blur stems from the response time, which is the duration taking a pixel to change its intensity, that despite its reduction along the years still introducing some amount of blur (see, e.g., \cite{rtingsMotionBlurofTVs}).
As in \cite{klompenhouwer2004motion,har2008lcd,chan2011lcd}, we consider motion blur arising only from the hold-type method of the LCD. 
The reader is referred to \cite{klompenhouwer2004motion,har2008lcd,chan2011lcd,pan2005lcd,masia2013survey} for additional discussions on the above described CRT and LCD motion artifacts.
In this section we will rely on existing models and problem settings addressing the LCD motion blur, and utilize our method from Section \ref{sec:The Proposed Method} for adjusting HEVC video coding to pre-compensate the perceived motion blur.

The two prominent signal-processing models of LCD motion blur were developed in the frequency \cite{klompenhouwer2004motion} and signal \cite{pan2005lcd} domains, considering the display impulse response and the human visual system (HVS) mechanisms of motion tracking and spatio-temporal low-pass filtering. A later model \cite{chan2011lcd} interpreted the former analyses to the case of discrete video signals, and approximated the temporal blur operator as an intra-frame spatial degradation determined by the current motion. This spatio-temporal equivalence of motion-blur degradation due to the hold-type nature of the LCD was used in various forms in \cite{kurita2001Moving,klompenhouwer2005temporal,klompenhouwer2006comparison,tourancheau2009lcd}. We here follow the model for LCD motion-blur given in \cite{chan2011lcd} to be aligned with the problem settings defined therein. 

The $ k^{th} $ frame of the displayed video is a $ W \times H $ two-dimensional discrete signal, comprised of $ N_f = W\cdot H $ pixels that their column-stack form is denoted here as $ \vec{v}_k \in \mathbb{R}^{N_f} $. The perceived image corresponding to the $ k^{th} $ frame is 
\begin{IEEEeqnarray}{rCl}
	\label{eq:LCD Motion Blur - Model - perceived degraded frame}
	\tilde{\vec{v}}_k = \mtx{H}_k \vec{v}_k
\end{IEEEeqnarray}
where $ \mtx{H}_k $ is a $ N_f \times N_f $ matrix representing the motion-blur as a spatial operator. The $ r^{th} $ row of $ \mtx{H}_k $ specifies the blur operation producing the $ r^{th} $ pixel of the degraded frame. The local blur operation is determined by the associated motion vector that may vary for different pixels.
For example, assume the $ r^{th} $ pixel corresponds to the motion vector $ \left(0,-3\right) $ describing a vertical motion upwards in 3 pixels with respect to the previous frame, accordingly, (assuming $ r $ corresponds to a coordinate sufficiently distant from the frame boundaries in its 2D arrangement) the $ r^{th} $ row of $ \mtx{H}_k $ should be formed from the following entries 
\begin{IEEEeqnarray}{rCl}
	\label{eq:LCD Motion Blur - blur matrix row example}
	 \mtx{H}_k \left[r,c\right]  = \left\{ {\begin{array}{*{20}{c}}
	 		\frac{1}{3} & {,\text{for~~} c = r, r-1, r-2 } \\ 
	 		0&{,\text{otherwise}. } 
	 	\end{array}} \right.  
\end{IEEEeqnarray}
A detailed numerical method for defining the blur kernel given a motion vector was described in \cite{chan2011lcd}. An important particular case occurs when the frame motion is global,  leading to a block-circulant matrix $\mtx{H}_k$.

\subsection{The Proposed Method for Motion-Blur Reduction}
\label{subsec:The Proposed Method for Motion-Blur Reduction}

We now turn to translate our general method given in Section \ref{sec:The Proposed Method} to the specific degradation of LCD motion blur. 
The considered video signal is a sequence of $ T $ frames, each of $ W\times H $ pixels. The column-stack form of the video is denoted as $ \vec{x} \in \mathbb{R}^{N} $ where $ N = T \cdot W \cdot H $ is the total amount of pixels. The signal $ \vec{x} $ is comprised from a concatenation of the $ T $ frames, i.e., 
\begin{IEEEeqnarray}{rCl}
	\label{eq:Motion-Blur - video signal as a concatenation of frames}
	{\vec{x}} = \left[ {\begin{array}{*{20}{c}}
			{{\vec{x}_{1}}} \\ 
			\vdots  \\ 
			{{\vec{x}_{T}}} 
		\end{array}} \right]
\end{IEEEeqnarray}
where $ {\vec{x}_{i}} \in \mathbb{R}^{N_f} $ is the column-stack form of the $ i^{th} $ frame, and $N_f = W\cdot H$ is the number of pixels in a frame.

The degradation considered here is modeled as independent spatial operations on each of the frames. Accordingly, the full signal degradation operator is the following block-diagonal matrix:
\begin{IEEEeqnarray}{rCl}
	\label{eq:Motion-Blur - block-diagonal form of H}
	\mtx{H}  = {\begin{bmatrix}
		\mtx{H}_1  & \vec{0} & \cdots & \vec{0}      \\[0.3em]
		\vec{0} & \mtx{H}_2 & \vec{0} & \vdots      \\[0.3em]
		\vdots & \vec{0} & \ddots & \vec{0}      \\[0.3em]
		\vec{0} & \cdots & \vec{0} & \mtx{H}_T     \\[0.3em]
	\end{bmatrix}}
\end{IEEEeqnarray}
where the $ N_f \times N_f $ matrix $ \mtx{H}_i $ is the spatial blur operator of the $ i^{th} $ frame.

Then, the block-diagonal structure (\ref{eq:Motion-Blur - block-diagonal form of H}) lets us to decompose the optimization (\ref{eq:degradation-oriented rate-distortion optimization - iterative solution - inversion}), which is an intermediate stage in the ADMM iteration, into the following frame-level optimizations: 
\begin{IEEEeqnarray}{rCl}
	\label{eq:Motion-Blur - degradation-oriented rate-distortion optimization - iterative solution - inversion - frame level}
	\hat{ \vec{z}}^{(t)}_i = \mathop {\text{argmin}}\limits_{{\vec{z}_i}\in\mathbb{R}^{N_f} } \left\| { \vec{x}_i - \mtx{H}_i \vec{z}_i } \right\|_2^2 + \frac{\beta}{2}{\left\| {  \vec{z}_i - \tilde{ \vec{v}}^{(t)}_i } \right\|_2^2} ~~,i=1,...,T \nonumber\\
\end{IEEEeqnarray}
where $ \hat{ \vec{z}}^{(t)}_i $ and $ {\tilde{ \vec{v}}^{(t)}_i } $ are the  column-vector forms of the $ i^{th} $ frames of the video signals $ \hat{ \vec{z}}^{(t)} $ and $ \tilde{ \vec{v}}^{(t)} $, respectively.
The analytic solution of the $ i^{th} $ frame optimization from (\ref{eq:Motion-Blur - degradation-oriented rate-distortion optimization - iterative solution - inversion - frame level}) is 
\begin{IEEEeqnarray}{rCl}
	\label{eq:Motion-Blur - degradation-oriented rate-distortion optimization - iterative solution - inversion - analytic form - frame level}
	\hat{ \vec{z}}^{(t)}_i = \left(  \mtx{H}^{T}_i\mtx{H}_i + \frac{\beta}{2} \mtx{I}  \right)^{-1} \left( \mtx{H}^T_i \vec{x}_i + \frac{\beta}{2} \tilde{ \vec{v}}^{(t)}_i  \right).
\end{IEEEeqnarray}
This computationally important update of Algorithm \ref{Algorithm:Proposed Method} is employed in Algorithm \ref{Algorithm:Proposed Method - LCD motion blur} describing the video compression method compensating a post-decompression LCD motion blur.
\begin{algorithm}
	\caption{Proposed Method: Video Coding Adjusted to Compensate LCD Motion Blur}
	\label{Algorithm:Proposed Method - LCD motion blur}
	\begin{algorithmic}[1]
		\State Inputs: $ \vec{x} $, $ \theta $, $ \beta $.
		
		\State Set frame blur operators $\left\lbrace \mtx{H}_i \right\rbrace_{i=1}^T$ based on motion estimation.
		
		\State  Initialize $ {\hat{\vec{z}}}^{(0)} = \vec{x} $ , $\vec{u}^{(1)} = \vec{0}$, $t = 1$.
		\Repeat 
		
		\State $ \tilde{ \vec{z}}^{(t)} = \hat{\vec{z}}^{(t-1)} - \vec{u}^{(t)} $
		\State $ \hat{\vec{v}}^{(t)} = {CompressDecompress}_{\theta}\left( \tilde{\vec{z}}^{(t)} \right). $
		\vspace{0.05in}
		
		\State $ \tilde{ \vec{v}}^{(t)} = \hat{\vec{v}}^{(t)} + \vec{u}^{(t)} $
		\State Form $\hat{ \vec{z}}^{(t)}$ via frame-level solutions for $i=1,...,T$ : 
		$~~~~~~~\hat{ \vec{z}}^{(t)}_i = \left(  \mtx{H}^{T}_i\mtx{H}_i + \frac{\beta}{2} \mtx{I}  \right)^{-1} \left( \mtx{H}^T_i \vec{x}_i + \frac{\beta}{2} \tilde{ \vec{v}}^{(t)}_i  \right)$
		
		\State $\vec{u}^{(t+1)} = \vec{u}^{(t)} + \left( \hat{ \vec{v}}^{(t)} - \hat{\vec{z}}^{(t)} \right)$
		
		\State $ t \gets t + 1$
		\Until{stopping criterion is satisfied}
		\State Output: Binary compressed data obtained in the last application of Stage 6.
	\end{algorithmic}
\end{algorithm}

\subsection{Experimental Results}
\label{subsec:Motion-Blur Reduction Experimental Results}

We evaluated our method by adjusting the HEVC video coding standard to compensate the perceptual motion-blur caused by LCD devices. As in previous works for LCD motion-blur reduction \cite{har2008lcd,chan2011lcd} we considered the 'Shields' and 'Stockholm' sequences (60 frames per second) \cite{video_test_media}, having global horizontal camera motions of -3 pixels/frame and 2 pixels/frame, respectively. The considered video segments were defined as 120 frames of 480$\times$480 pixels taken from the 720p sequences mentioned above.

\begin{figure*}[]
	\centering
	{{\subfloat[]{\label{fig:shields_psnr_curves}\includegraphics[width=0.35\textwidth]{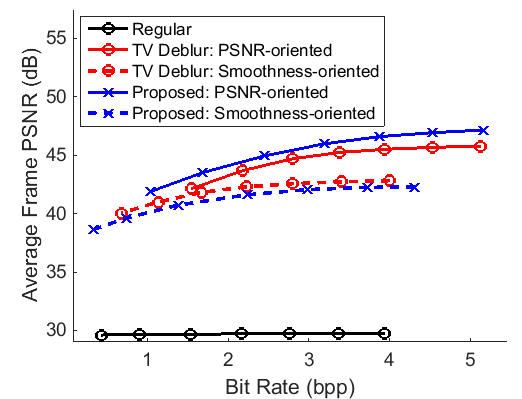}}}
		{\subfloat[]{\label{fig:shields_ssim_curves}\includegraphics[width=0.35\textwidth]{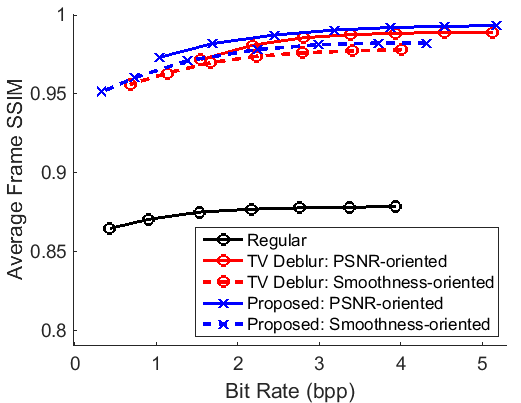}}}}
	\caption{LCD motion blur experiment for the 'Shields' sequence. The performance of our method in its PSNR-oriented (solid blue line) and smoothness-oriented (dashed blue line) modes is compared to preprocessing using total-variation deblurring \cite{chan2011augmented} in its PSNR-oriented (solid red line) and smoothness-oriented (dashed red line) parameter settings, and to a regular compression-decompression procedure without any additional processing (solid black line). The average frame-PSNR and average frame-SSIM are evaluated in (a) and (b), respectively.} 
	\label{Fig:Experiments - LCD motion blur - shields - performance curves}
\end{figure*}

\begin{figure*}[]
	\centering
	{{\subfloat[]{\label{fig:stockholm_psnr_curves}\includegraphics[width=0.35\textwidth]{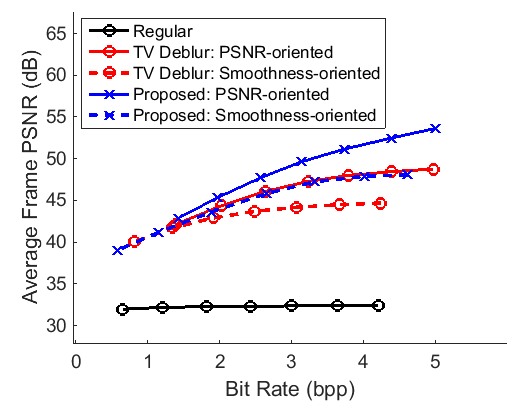}}}
		{\subfloat[]{\label{fig:stockholm_ssim_curves}\includegraphics[width=0.35\textwidth]{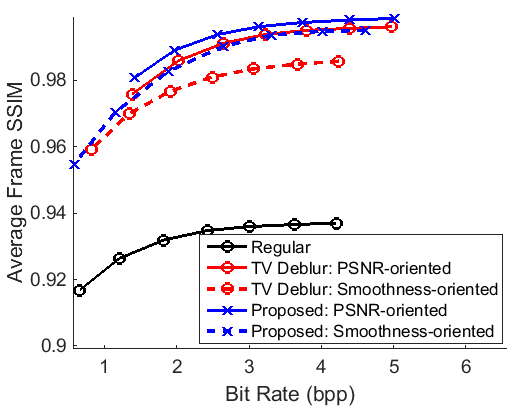}}}}
	\caption{LCD motion blur experiment for the 'Stockholm' sequence. The performance of our method in its PSNR-oriented (solid blue line) and smoothness-oriented (dashed blue line) modes is compared to preprocessing using total-variation deblurring \cite{chan2011augmented} in its PSNR-oriented (solid red line) and smoothness-oriented (dashed red line) parameter settings, and to a regular compression-decompression procedure without any additional processing (solid black line). The average frame-PSNR and average frame-SSIM are evaluated in (a) and (b), respectively.} 
	\label{Fig:Experiments - LCD motion blur - stockholm - performance curves}
\end{figure*}

In the experiments of this section we use the HEVC implementation given in the reference software HM 15.0 \cite{hevc_reference_software} set to the 'random access' profile, where powerful motion-compensation procedures together with P and B frame types are employed. 
The presented comparison (Figs. \ref{Fig:Experiments - LCD motion blur - shields - performance curves}-\ref{Fig:Experiments - LCD motion blur - stockholm - performance curves}) consider HEVC compression operated for QP values between 1 and 19 in jumps of 3. 
The performance evaluations in Figs. \ref{fig:shields_psnr_curves},\ref{fig:stockholm_psnr_curves} rely on the
average PSNR of the frames. The comparisons in Figs. \ref{fig:shields_ssim_curves},\ref{fig:stockholm_ssim_curves}
consider the value of the SSIM metric \cite{wang2004image} averaged over all the frames considered.
The basic reference performance is the regular compression where no pre or post processing is done and, consequently, the respective performance is inferior to the other processing-aided methods (see black curves in Figs. \ref{Fig:Experiments - LCD motion blur - shields - performance curves}-\ref{Fig:Experiments - LCD motion blur - stockholm - performance curves}).

The main competing approach is to precede the compression with a video deblurring method addressing the motion blur using spatio-temporal total-variation (TV) regularization \cite{chan2011augmented} (note that the deblurring technique in \cite{chan2011augmented} extends and improves upon the LCD motion-blur reduction method in \cite{chan2011lcd}). Importantly, the methods in \cite{chan2011lcd,chan2011augmented} consider the video motion deblurring without any aspect of compression, thus, we employ them in the problem settings considered here as a pre-compression stage. Accordingly, we optimized the parameters of \cite{chan2011augmented} to provide a high average frame-PSNR in our settings (see red solid-line performance curves in Figs. \ref{Fig:Experiments - LCD motion blur - shields - performance curves}-\ref{Fig:Experiments - LCD motion blur - stockholm - performance curves}).
In addition, high PSNR is not necessarily coupled with high visual quality, as the perceived video may look noisy and/or flickery (see, e.g., \cite{chan2011lcd}). Therefore, we also define an additional parameter setting of \cite{chan2011augmented} to provide more visually pleasing results at the expense of the PSNR (the performance curves of this smoothness-oriented setting appear as red dashed-line curves in Figs. \ref{Fig:Experiments - LCD motion blur - shields - performance curves}-\ref{Fig:Experiments - LCD motion blur - stockholm - performance curves}. 
Moreover, we noticed that the total-variation deblurring method produces artifacts along the vertical borders of the frames (the horizontal borders are artifact free because the global motion is horizontal). Accordingly, we gave the total-variation deblurring method a larger portion of the frame with margins of 100 pixels in each side, then, these margins are removed before given to the compression. This procedure was carried out only for the total-variation deblurring approach.

We evaluated our method in two modes: the first aims to a high PSNR by the setting $ \beta = 10 \tilde{\beta} $ where $ \tilde{\beta} $ is defined by the QP-dependent rules in (\ref{eq:beta in image experiments}).
The second version, referred to as smoothness-oriented, is determined by setting $ \beta = 50 \tilde{\beta} $ that leads to an increased spatio-temporal smoothness (a visually pleasing property) at the expense of the PSNR.
Both of these settings employed a maximal number of 10 iterations, or stopped earlier if convergence or divergence are detected. The rules defining convergence/divergence are as in the image experiment presented in Section \ref{subsec: Image experiments - our method}, but with the different threshold values of $ 0.5 \cdot T $ and $ 50 / T $ for convergence and divergence, respectively (also note the dependency on the number of frames $ T $). 
The performance curves of the PSNR-oriented and the smoothness-oriented modes of our method appear in Figs. \ref{Fig:Experiments - LCD motion blur - shields - performance curves}-\ref{Fig:Experiments - LCD motion blur - stockholm - performance curves} as blue solid-lines and blue dashed-lines, respectively. 

Figures \ref{Fig:Experiments - LCD motion blur - shields - performance curves}-\ref{Fig:Experiments - LCD motion blur - stockholm - performance curves} show that our method greatly outperforms the regular compression procedure: a BD-PSNR gain of 13.90 dB was achieved for the 'Shields' sequence (Fig. \ref{fig:shields_psnr_curves}), and a gain of 13.28 dB was obtained for the 'Stockholm' sequence (Fig. \ref{fig:stockholm_psnr_curves}).
Figures \ref{fig:shields_ssim_curves},\ref{fig:stockholm_ssim_curves} present also significant gains in SSIM terms. 

Let us examine the performance of our approach with respect to the total-variation deblurring technique, both in their PSNR-oriented settings (the blue and red solid-line curves in Figs. \ref{Fig:Experiments - LCD motion blur - shields - performance curves}-\ref{Fig:Experiments - LCD motion blur - stockholm - performance curves}). Considering the 'Shields' (Fig. \ref{fig:shields_psnr_curves}) and 'Stockholm' (Fig. \ref{fig:stockholm_psnr_curves}) sequences, our method achieved respective BD-PSNR gains of 1.06 dB and 2.16 dB over the total-variation deblurring technique. Figures \ref{fig:shields_ssim_curves},\ref{fig:stockholm_ssim_curves} exhibit that our PSNR-oriented method is better than the TV deblurring approach also with respect to the SSIM quality metric.

The third comparison considers our method with respect to the total-variation deblurring approach, both in their smoothness-oriented settings (the blue and red dashed-line curves in Figs. \ref{Fig:Experiments - LCD motion blur - shields - performance curves}-\ref{Fig:Experiments - LCD motion blur - stockholm - performance curves}). 
Considering the 'Shields' (Fig. \ref{fig:shields_psnr_curves}) and 'Stockholm' (Fig. \ref{fig:stockholm_psnr_curves}) sequences, our method differs from the TV-deblurring technique in BD-PSNR values of $-0.62$ dB and $0.98$ dB. One should recall that this is the smoothness-oriented settings, thus, visual quality is preferred over optimizing the PSNR. Indeed, examining the SSIM-bitrate curves of the 'Shields' (Fig. \ref{fig:shields_ssim_curves}) and 'Stockholm' (Fig. \ref{fig:stockholm_ssim_curves}) sequences, exhibit that our smoothness-oriented method obtains the respective average SSIM gains of $ 3.4 \times 10^{-3}$ and $ 5.8 \times 10^{-3}$ over the smoothness-oriented TV deblurring technique. These results point on the good visual quality offered by the smoothness-oriented settings of our method.

%The curves of the PSNR-oriented methods in Figs. \ref{Fig:Experiments - LCD motion blur - shields - performance curves}-\ref{Fig:Experiments - LCD motion blur - stockholm - performance curves} reflect the significant gains in PSNR and SSIM achieved by our method with respect to the total-variation deblurring. 
%Specifically, calculating the BD-PSNR metric \cite{bjontegaard2001calculation,BDPSNR_Matlab}, measuring the average distance between the performance curves, showed that for the 'Stockholm' sequence (Fig. \ref{fig:stockholm_psnr_curves}) our PSNR-oriented method achieved a BD-PSNR gain of 13.28 dB and 2.16 dB with respect to the regular compression and to the PSNR-oriented total-variation deblurring technique. Moreover, for the 'Shields' sequence our PSNR-oriented method achieved a BD-PSNR gain of 13.90 dB and 1.06 dB with respect to the regular compression and to the PSNR-oriented total-variation deblurring technique.


\begin{figure*}[]
	\centering
	{\subfloat[Original]{\label{fig:shields_frame100_input}\includegraphics[width=0.19\textwidth]{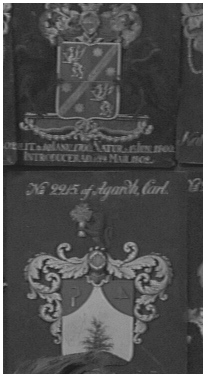}}}~~
	\\
	{{\subfloat[Regular: Displayed \newline 3.940 bpp]{\label{fig:shields_frame100_regular_displayed}\includegraphics[width=0.19\textwidth]{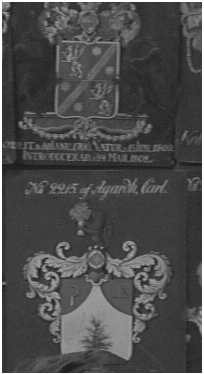}}}~~
	{\subfloat[TV Smoothness-Oriented: Displayed \newline 3.411 bpp]{\label{fig:shields_frame100_TV_smoothness_oriented_displayed}\includegraphics[width=0.19\textwidth]{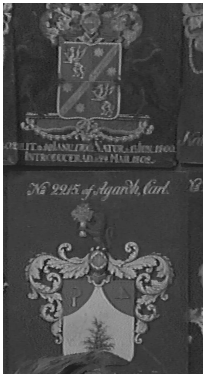}}}~~
	{\subfloat[TV PSNR-Oriented: Displayed \newline 3.377 bpp]{\label{fig:shields_frame100_TV_psnr_oriented_displayed}\includegraphics[width=0.19\textwidth]{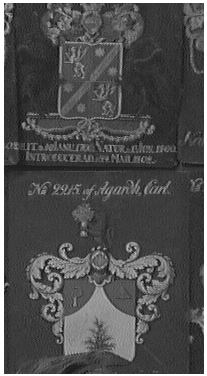}}}~~
	{\subfloat[Proposed Smoothness-Oriented: Displayed \newline 3.248 bpp]{\label{fig:shields_frame100_our_smoothness_oriented_displayed}\includegraphics[width=0.19\textwidth]{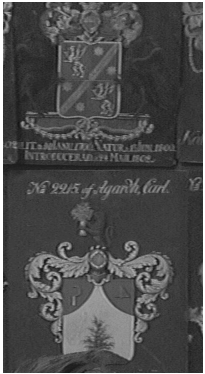}}}~~
	{\subfloat[Proposed PSNR-Oriented: Displayed \newline 3.192 bpp]{\label{fig:shields_frame100_our_psnr_oriented_displayed}\includegraphics[width=0.19\textwidth]{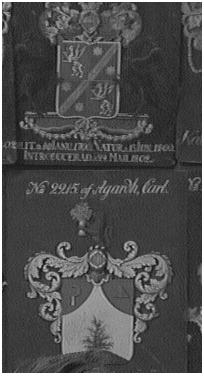}}}
	}
	\\
	{{\subfloat[Regular: Perceived \newline PSNR 30.72 dB \newline SSIM 0.8884]{\label{fig:shields_frame100_regular_perceived}\includegraphics[width=0.19\textwidth]{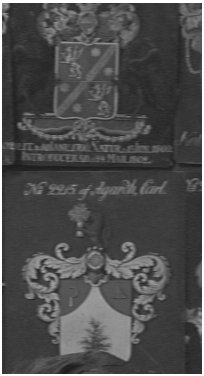}}}~~
	{\subfloat[TV Smoothness-Oriented: Perceived \newline PSNR 43.39 dB \newline SSIM 0.9796]{\label{fig:shields_frame100_TV_smoothness_oriented_perceived}\includegraphics[width=0.19\textwidth]{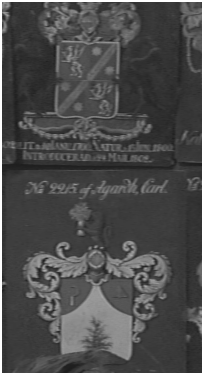}}}~~
	{\subfloat[TV PSNR-Oriented: Perceived \newline PSNR 45.49 dB \newline SSIM 0.9873]{\label{fig:shields_frame100_TV_psnr_oriented_perceived}\includegraphics[width=0.19\textwidth]{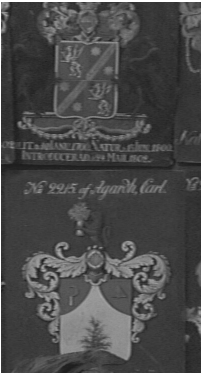}}}~~
	{\subfloat[Proposed Smoothness-Oriented: Perceived \newline PSNR 42.99 dB \newline SSIM  0.9820]{\label{fig:shields_frame100_our_smoothness_oriented_perceived}\includegraphics[width=0.19\textwidth]{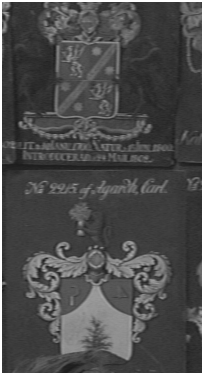}}}~~
	{\subfloat[Proposed PSNR-Oriented: Perceived \newline \bf{PSNR 46.02 dB} \newline \bf{SSIM 0.9894}]{\label{fig:shields_frame100_our_psnr_oriented_perceived}\includegraphics[width=0.19\textwidth]{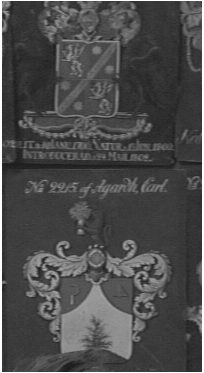}}}
	}
	\caption{LCD motion blur experiment for the sequence 'Shields' (grayscale, 120 frames at 60fps, 480x480 pixels) where a global horizontal motion of -3 pixels/frame causes the blur. An exemplary frame segment is presented. (a) the original frame segment. 
	(b)-(f) are the displayed frame segments using each of the five examined methods.
	(g)-(k) are the simulated perceived frame segments corresponding to each of the displayed frames. The presented PSNR and SSIM evaluations are for the complete perceived frame, and the bit-rates are those measured for the compression of the complete sequence of 120 frames.} 
	\label{Fig:Experiments - LCD motion blur - shields}
\end{figure*}

\begin{figure*}[]
	\centering
	{{\subfloat[Regular]{\label{fig:shields_frame100_regular_perceived_difference}\includegraphics[width=0.19\textwidth]{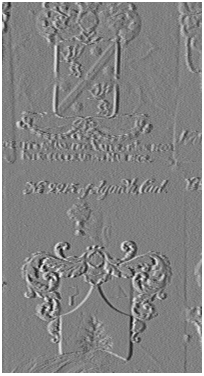}}}~~{\subfloat[TV 								Smoothness-Oriented]{\label{fig:shields_frame100_TV_smoothness_oriented_perceived_difference}\includegraphics[width=0.19\textwidth]{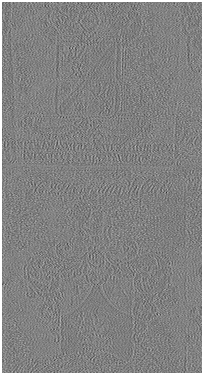}}}~~
	{\subfloat[TV PSNR-Oriented]{\label{fig:shields_frame100_TV_psnr_oriented_perceived_difference}\includegraphics[width=0.19\textwidth]{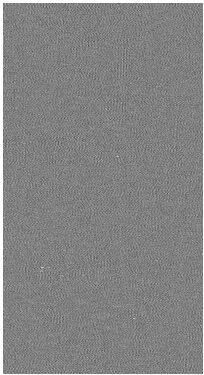}}}~~
	{\subfloat[Proposed Smoothness-Oriented]{\label{fig:shields_frame100_our_smoothness_oriented_perceived_difference}\includegraphics[width=0.19\textwidth]{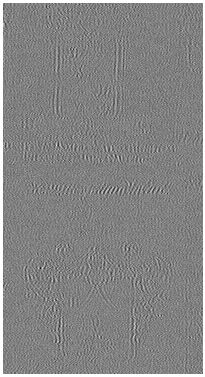}}}~~
	{\subfloat[Proposed PSNR-Oriented]{\label{fig:shields_frame100_our_psnr_oriented_perceived_difference}\includegraphics[width=0.19\textwidth]{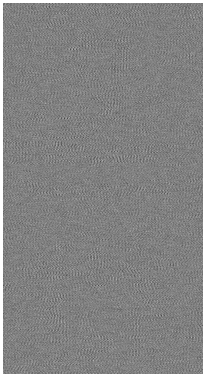}}}
	}
	\caption{LCD motion blur experiment for the sequence 'Shields'. Here we present the difference images between the perceived and the original frame segments, for each of the methods presented in Fig. \ref{Fig:Experiments - LCD motion blur - shields}:
		(a) regular, (b) TV smoothness-oriented, (c) TV PSNR-oriented, (d) proposed smoothness-oriented, and (e) proposed PSNR-oriented. 
		The difference images are presented in grayscale by scaling the value range of $ \left[ -50, 50 \right] $ for the regular approach, and the value range of $ \left[-15, 15\right] $ for the other methods.}
	\label{Fig:Experiments - LCD motion blur - shields - difference frames}
\end{figure*}

In Figure \ref{Fig:Experiments - LCD motion blur - shields} we provide a visual demonstration of the results obtained for a frame segment from the sequence 'Shields' (Fig. \ref{fig:shields_frame100_input}).
As the various methods do not produce equal bit-rates, the comparison is for relatively close bit-rates -- specifically, the results presented for our method obtained using lower bit-rates than the other techniques. 
Figures \ref{fig:shields_frame100_regular_displayed}-\ref{fig:shields_frame100_our_psnr_oriented_displayed} show the frames given to the display using the various methods and their settings. Our method (Fig. \ref{fig:shields_frame100_our_smoothness_oriented_displayed}-\ref{fig:shields_frame100_our_psnr_oriented_displayed}) as well as the other deblurring-based approach (Fig. \ref{fig:shields_frame100_TV_smoothness_oriented_displayed}-\ref{fig:shields_frame100_TV_psnr_oriented_displayed}) provide sharpened images to display. Figures \ref{fig:shields_frame100_regular_perceived}-\ref{fig:shields_frame100_our_psnr_oriented_perceived} exhibit the simulated perceived image (i.e., the displayed frame after the blur degradation modeled above). Evidently, the perceived images corresponding to the pre-compensating techniques highly resemble the original frame.
Moreover, Fig. \ref{Fig:Experiments - LCD motion blur - shields - difference frames} shows for each of the methods the difference between the simulated-perceived and the original images, suggesting that our PSNR-oriented method avoids noticeable image-detail loss, at the expense of some textural-noise that can be attenuated using the smoothness-oriented settings.
Importantly, the PSNR-oriented settings of our method achieve the highest PSNR and SSIM values for the presented frame, while using a lower bit-rate than the regular and the total-variation approaches.

\section{Conclusion}
\label{sec:Conclusion}
In this paper we studied an image/video compression problem where a linear degradation affects the decompressed signal. We handled the difficult rate-distortion optimization using the ADMM approach, resulting in an iterative procedure relying on a standard compression technique. 
Our method was shown to be effective in experiments of adjusting HEVC's image and video compression to post-decompression blur degradations.
We consider the proposed method as a promising paradigm for addressing complicated rate-distortion optimizations arising in intricate systems and applications. 
Future work may extend our approach for optimizing systems where several display types are possible. Another interesting process to examine includes a degradation that precedes the compression (e.g., an acquisition stage) in addition to the post-decompression degradation (e.g., rendering).
Other intriguing research directions may explore the utilization of our idea for optimizing compression with respect to various perceptual distortion metrics or to nonlinear systems.

%\section*{Acknowledgment}

\ifCLASSOPTIONcaptionsoff
  \newpage
\fi



\bibliographystyle{IEEEtran}
\bibliography{IEEEabrv,compression_with_degradation_journal__refs}

%\begin{thebibliography}{1}
%\end{thebibliography}

%\begin{biography}[{\includegraphics[width=1in,height=1.25in,clip,keepaspectratio]{mshell}}]{Michael Shell}

%\newpage


%\vfill


\end{document}